# Model Uncertainty under Non-Gaussian Errors: Bayesian Model Averaging and Selection in Stochastic Frontier Models

Kamil Makieła

Krakow University of Economics
Department of Econometrics and Operations Research
kamil.makiela@uek.krakow.pl

## Abstract

The paper investigates Bayesian Model Averaging and Selection (BMA/S) under non-standard stochastic assumptions, focusing on stochastic frontier analysis (SFA). We propose fast, reliable procedures for inference in the normal-exponential stochastic frontier model and examine whether accounting for asymmetric disturbances affects model averaging and/or selection outcomes relative to the conventional Gaussian-error BMA/S. Particular attention is given to moderate-dimensional covariate selection problems typical in SFA applications. We demonstrate that, with appropriate search strategies and parallelization techniques, exhaustive model search can be computationally feasible and, in some cases, more practical than stochastic search alternatives. A Monte Carlo simulation study is used to compare the proposed SF-BMA/S procedure with standard Gaussian-error BMA/S under varying levels of inefficiency-to-noise ratio and signal strength with respect to the data generating process. The results show that accounting for stochastic frontier structures may affect posterior inference and model averaging outcomes, especially in scenarios where efficiency analysis is most sensible.

**Funding:** the publication was financed from the subsidy granted to the Krakow University of Economics, project number: 056/EIE/2026/POT.

## 1 Introduction

As computational power increases and variable-rich datasets become more available, research strategies for building regression models shift towards automated procedures. Probably the most renowned method for model search used in economics is Bayesian Model Averaging or Selection; BMA/S hereafter (Steel, 2020). Using BMA/S, one may either pool inference about quantities common to all models (model averaging) or select the best model given the data (model selection). While the method appears to have recently gained popularity in machine-learning (Li *et al.*, 2024), it has been well-known in econometrics for decades. Early discussions about BMA/S can be traced to Zellner (1971), Mitchell and Beauchamp (1988), Osiewalski and Steel (1993) and Raftery et al. (1997).

Running a full, non-stochastic BMA/S can still pose a computational challenge as it requires 'looking through' all $2^k$ possible models, where $k$ is the number of candidate variables (Fernández *et al.*, 2001; Hoeting *et al.*, 1999). Thus, such full BMA/S implementations, often called exhaustive search or brute-force search, are rare. Applications usually rely on stochastic BMA/S procedures, partially due to a well-known package available in R (Raftery *et al.*, 2005). Such procedures involve using Markov Chain Monte Carlo Model Comparison (MC3) techniques. However, these require fine-tuning the simulation's parameters to achieve appropriate acceptance/jump rates for the sampler (Fernández *et al.*, 2001). Regardless of the search algorithm it is important to note that the model structure under investigation in BMA/S is usually simplistic and motivated by computational convenience rather than economic correctness (Ley and Steel, 2009), with a few recent exceptions to this rule (Steiner and Steel, 2026; Tasiopoulos *et al.*, 2026).

Despite the increasing popularity of BMA/S-like procedures, a notable research gap remains regarding their use under non-standard stochastic assumptions, such as Stochastic Frontier Analysis (SFA). This is somewhat surprising for at least three reasons. First, efficiency scores in SF models are contingent on the parametric specification of the frontier function. Thus, choosing the optimal specification given data (understood, e.g., in terms of covariate selection) seems particularly important, given the main goal of SFA, which is efficiency analysis. Second, moderate-dimensional covariate selection problems involving approximately up to $k$=20 candidate regressors remain standard in many econometric applications and methodological studies (George and McCulloch, 1993; Miller, 2002; Smith and Kohn, 1996). This is even more typical in SFA, where empirical specifications are relatively parsimonious and usually theory-driven (Makieła et al., 2025). For example, in a translog function, which is widely used in SFA as a generalization of a standard Cobb-Douglas production/cost function, the first-order approximation terms (i.e., the Cobb-Douglas part) may be retained in the final model regardless of their contribution to the overall fit as dictated by the underlying economic theory. This, in turn, substantially reduces the total number of models that need to be considered. Third, advances in parallel computing have made exhaustive model search or near-exhaustive model search substantially more feasible. Despite these developments, model uncertainty, understood in terms of covariate selection, is still commonly addressed in BMA by methods that rely on classic chained-linked stochastic search and rather simplistic assumptions about the error term (Fernández *et al.*, 2001). In this paper, we demonstrate that exhaustive search, when combined with an effective search strategy and appropriate parallelization techniques, can be computationally feasible for moderate-dimensional covariate selection problems (e.g., $k$<30). Moreover, in some cases it may prove more practical than its stochastic search counterparts (e.g., $k$<25).

By implementing fast, reliable BMA/S procedures for SFA, we want to learn if accounting for non-Gaussian error structures, like the asymmetric disturbances traditional to SF models, impacts Bayesian model selection and averaging outcomes relative to the conventional Gaussian-error models. And if so, how relevant these changes are, and under what circumstances they matter most. As mentioned, proper covariate selection seems particularly important in SFA because efficiency scores are contingent on the parametric specification of the frontier function. Hence, there is a straightforward motivation to use BMA/S techniques in SFA whenever the theoretical framework is scarce or ambiguous.

In this paper, we focus on the Bayesian normal-exponential SF model for two reasons. First, because exponential inefficiency terms, as opposed to half-normal and, more importantly, truncated-normal, may provide better identification of the error components, which is crucial for a fast, stable SFA procedure. Second, the priors for the normal-exponential SF model proposed by van den Broeck et al. (1994), which we also use here, are less informative compared to, e.g., normal-half-normal (Makieła and Mazur, 2020). These two aspects have made the normal-exponential model particularly appealing when considering our choices for Bayesian averaging/selection of stochastic frontier models; SF-BMA/S hereafter.

The paper is structured as follows. Section 2 discusses Bayesian model averaging as well as the Bayesian normal-exponential SF model used in the study. It then moves on to discuss bottlenecks in implementing Bayesian model averaging and selection for stochastic frontier models, along with strategies on how they can be handled. Section 3 presents a Monte Carlo simulation study, which was used to assess the performance of the proposed approach relative to standard Gaussian-error BMA/S. Section 4 shows empirical examples, and Section 5 concludes with a discussion.

# 2 Methodology

## 2.1 Key aspects of Bayesian model averaging and selection

Let us consider $m$ statistical models defined on the same observation space $Y$:

$$M_j: p_j(y|\theta_j) = p_j(y|\alpha, \omega_j), \quad j = 1, \dots, m, \tag{1}$$

where $y \in Y$ is the matrix (vector) of observations being modelled. The vector of unobserved quantities $\theta_j = (\alpha, \omega_j) \in \Theta_j = \Lambda \times H_j$ consists of two vectors: $\alpha$, which groups quantities that are common to all models (i.e., inefficiency terms, intercept, etc.), and $\omega_j$, which groups model specific quantities. Formal model comparison in the Bayesian framework is based on posterior probabilities; see, e.g., Osiewalski and Steel (1993). To compute these probabilities, the well-known Bayes' theorem is used:

$$p(M_j|y) = \frac{p(M_j)p(y|M_j)}{\sum_{k=1}^{m} p(M_k)p(y|M_k)}, \tag{2}$$

where

$$p(y|M_j) = p_j(y) = \int_{\Theta_j} p_j(y|\theta_j)p_j(\theta_j)d\theta_j, \quad j = 1, \dots, m, \tag{3}$$

is the integrated likelihood of the observation vector in model $M_j$, and $p(M_j)$ is the prior probability of the model. For simplicity, we use equal prior odds for all possible model specifications throughout the paper, though we note that there are other strategies available in the literature (Makieła and Osiewalski, 2018).

The posterior probability of model $M_j$ is determined by the product of its prior probability and the value of the integrated likelihood, which constitutes a natural Bayesian measure of model fit. Based on posterior model probabilities, one may either combine (pool, average) inference about quantities common to all models or select the model with the highest posterior probability (i.e., the best model). The former is usually known as Bayesian inference pooling or Bayesian model averaging (BMA). The latter is referred to as Bayesian model selection and provides an intuitive, simplified solution when one model clearly dominates the posterior model probability ranking. We shall refer to them jointly as BMA/S. If the goal of BMA/S is covariate selection, then the total number of models to consider is $2^k$, where $k$ is the number of candidate-variables. In this paper we define the total number of all regression parameters in vector $\beta$, including the intercept, as $\kappa = k + k_0$ and set $k_0 = 1$, which indicates that we only leave out, quite obviously, the constant term from variable selection. Of course, since $k_0 \geq 1$ we can include any number of variables, apart from the constant, to be removed from the variable selection procedure, which is sometimes dictated by economics theory.

In contemporary applications, the parameter space is usually explored via Markov Chain Monte Carlo Model Comparison (MC3) methods, and thus a subset of relevant models is used in BMA/S. As already stated in the introduction, we focus on exact search strategies.

## 2.2 Stochastic frontier models

For our BMA/S procedure, we consider the following production-orientated pooled stochastic frontier normal-exponential model:

$$y_i = x_i\beta + v_i - u_i \quad (4)$$

where $y_i$ is the vector of observations on the dependent variable (e.g., production output), $x_i$ is the vector of explanatory variables (e.g., production inputs), $v_i$ is the normally distributed symmetric term with zero mean and unknown variance $\sigma_v$, $u_i$ is the nonnegative, exponentially distributed, inefficiency term with unknown mean and standard deviation $\sigma_u$, and $n$ is the number of observations. The Bayesian Common Efficiency Distribution (CED) normal-exponential SF specification is:

$$p(y, \beta, \sigma_v, \sigma_u, u|\text{data}) = p(\beta, \sigma_v, \sigma_u)\prod_{i=1}^{n}\left(f_g(u_i|1, \sigma_u^{-1})f_N(y_i|x_i\beta - u_t, \sigma_v^2)\right) \quad (5)$$

where $f_N(a, b)$ is the density function of the normal distribution with mean (a) and variance (b), $f_G(c, d)$ is the gamma density function with expected value $c/d$ and variance $c/d^2$; $p(\beta, \sigma_v, \sigma_u)$ is the prior distribution for $\beta$, $\sigma_v$ and $\sigma_u$. For $\sigma_u$, we assume inverse gamma $\sigma_u \sim IG(1, -\ln r^*)$, which is the equivalent of gamma prior on $\sigma_u^{-1}$, $p(\sigma_u^{-1}) = f_G(1, -\ln r^*)$; see van den Broeck et al. (1994). Hyperparameter $r^*$ is the prior median efficiency, which we set $r^* = 0.75$ throughout the paper; see, e.g., Makieła (2014) for a discussion on setting $r^*$. For $\sigma_v$, we consider the standard inverse gamma prior on its inversed square (precision): $\sigma_v^{-2} \sim IG(g_1/2\,, g_2/2\,)$ with $g_1 = \; g_2 = 10^{-4}$. For $\beta$, we use multivariate $k$-dimensional normal distribution $p(\beta) = f_N^{k+1}(b, C)$, where b is a zero-vector of length $k + 1$ (the intercept and $k$ regression coefficients), and C is a $(k + 1)\text{x}(k + 1)$ diagonal covariance matrix $C = Var(\beta) = 100I_{k+1}$. Thus, for the purpose of this paper $k$ is defined as the number of regressors excluding the constant. Lastly, we can simplify Bayesian SF model to standard Bayesian Gaussian-error linear regression:

$$p(y, \beta, \sigma_v|\text{data}) = p(\beta, \sigma_v)\prod_{i=1}^{n} f_N(y_i|x_i\beta, \sigma_v^2) \quad (6)$$

with the same priors on $\beta, \sigma_v$ to maintain prior coherence (Makieła and Mazur, 2020). Since this specification will be used also in a non-Bayesian setting (i.e., Maximum Likelihood estimation), we shall refer to it simply as the Gaussian-error model.

For our SF-BMA/S procedure, we need to efficiently evaluate the following two elements for every SF model: the statistical parameters (i.e., $\beta, \sigma_v, \sigma_u$) and the integrated likelihood. First, SF models do not have closed-form likelihoods, although they are sometimes described as such; or as 'approximately' closed-forms, see, e.g., Makieła and Mazur (2022), Tsionas (2012). For example, in normal-exponential SF model, once we integrate out laten inefficiency terms we get normal CDF in the likelihood, which is still an integral that needs to be evaluated numerically. Since fast algorithms for normal CDF evaluation are now standard in most computational packages, this is often misunderstood as a closed-form solution, which it is not. Following Makieła and Mazur (2022), we use a loglikelihood form marginalized with respect to inefficiency terms, which is a standard practice in classic, frequentist SFA; see e.g. Parmeter and Kumbhakar (2014). We use optimization algorithms available in MATLAB for fast evaluation of the likelihood for Maximum Likelihood estimation (ML hereafter) and the posterior for Bayesian estimation. Note, that efficiency of the optimization procedures is contingent on two elements: (i) starting points, which we take from OLS for ML, and then from ML for Bayesian estimation; and (ii) providing gradients, which are analytical in the normal-exponential case.

Furthermore, we use two model reparametrizations. Parametrization A, defined as $\theta = (\beta, \log\sigma_v\,, \log\sigma_u)$, is particularly useful for estimating integrated likelihood via Laplace approximation. This is because, as Makieła and Mazur (2022) point out, the posterior in this representation is more likely to resemble multivariate normal around its maximum compared to the original parametrization, and this is crucial for accurate Laplace approximation. Parametrization A, however, is not very effective for optimization algorithms because decomposition of the total (square root of) variance into the two components sometimes may not be well-identifiable and thus difficult to evaluate. Considering their logged forms helps, but it does not fully solve the problem. Ideally, optimization algorithms would like parameters to be orthogonal, and this is hardly the case in parametrization A. Hence, for numerical optimization purposes we use parametrization B, defined as $\theta = (\beta, \log \sigma_\varepsilon^2, \eta)$ where $\sigma_\varepsilon^2 = \sigma_v^2 + \sigma_u^2$, $\eta = \text{logit}(\gamma)$ and $\gamma = \frac{\sigma_u^2}{\sigma_v^2 + \sigma_u^2}$. This parametrization is based on Battese and Corra (1977) reparameterization of the variance components, which defines $\sigma_\varepsilon^2 = \sigma_u^2 + \sigma_v^2$ and $\gamma = \frac{\sigma_u^2}{\sigma_\varepsilon^2}$. This way we have two error parameters: (i) $\sigma_\varepsilon^2$, which accounts for the total variance and (ii) $\gamma$, which accounts for skewness of the compound error $\varepsilon$ with two extreme cases: $\gamma = 0$ if there is no inefficiency; $\gamma = 1$ if there is no symmetric disturbance. From the numerical stability perspective, we need to safeguard these two cases, which in the Bayesian estimation are largely handled by the prior. Also, for numerical stability, we estimate the parameters $(\sigma_\varepsilon^2, \gamma)$ in a transformed form $(\log \sigma_\varepsilon^2, \text{logit}\, \gamma)$. Although this parameterization does not invoke the aforementioned orthogonality, it provides sufficient identifiability and thus numerical stability. More notably, transformations of variables' densities from A to B and back, as well as to the original parametrization for which we defined priors, is analytical. No numerical approximations are required.

## 2.3 Bottlenecks in the SF-BMA/S procedure

We can identify two main bottlenecks for SF-BMA/S: accuracy and speed. First, we address the accuracy of ML and Bayesian estimation of a single SF model, having of course, speed considerations in mind. For ML estimation, we use model optimization tools that are readily available in MATLAB. To maximize loglikelihood (marginalized w.r.t. inefficiencies), we use parametrization B, because optimization algorithms work best when all parameters are properly identifiable (preferably even orthogonal). This may not necessarily be the case under parametrization A. Since speed is crucial, running even the simplest MCMC for Bayesian estimation is not feasible. Following Makieła and Mazur (2022), we turn to brute-force optimization algorithms to evaluate the maximum posterior, like the ones used for ML. We search for maximum under parameterization B using ML estimates as starting points. Also following Makieła and Mazur (2022), we compute the integrated likelihood using Laplace approximation under parametrization A. The posterior around its maximum in this parameterization is more symmetric and more likely to resemble multivariate normal than in B, especially for cases where $\gamma$ is close to zero or one (i.e., in such cases the posterior becomes substantially skewed in this dimension). Hence, the Laplace approximation under B can get distorted. As Makieła and Mazur (2022) show, Laplace approximation under parameterization A works surprisingly well even for much more complex SF models than normal-exponential and compared to more precise MC-based approaches (Pajor, 2017). Having BMA/S application in mind, this seems like a fair balance between speed and accuracy.

Probably the main bottleneck for the SF-BMA/S deployment is the substantial time needed for model evaluation. Even with the best coding practices in mind, one model evaluation takes at least a few milliseconds in MATLAB. Note that this is still quite fast, given that we need to move between parameterizations for estimation (B) and for the Laplace approximation of the integrated likelihood (A). The first, rather obvious, solution to the problem is parallelization. Since parallelization cannot be effectively implemented within MC3 algorithms, as they require chain-linked loops, we do not consider them and focus only on exhaustive search algorithms, i.e., full search of the entire parameter space ($2^k$). The upside of using exhaustive search is also that we do not need to worry about the usual pitfalls of MC3 (burn-in phase, the number of accepted draws, sampler's mixing properties etc.). This makes our approach more accessible to practitioners. The gain from parallelization in exhaustive search comes when we loop over model space because every model combination can be evaluated independently. So, this is *where* parallelization takes place in our procedure. It is easy to show that within certain limits of $k$ (up to about 12-14) this alone makes a huge difference. Also, it is important to note that the current development trend of computing infrastructure is heavily skewed towards parallel computing for AI-training purposes. So, the speed gain from such parallelization is likely to increase dramatically in the near future.

The second solution is to pre-screen Bayesian SF model space based on a faster, simpler method and retain only the potentially relevant set of models. This is because in practice an overwhelming number of possible variable combinations (i.e. models) is likely to be irrelevant, and thus it is warranted to use BMA/S only on models with nonnegligible posterior probability. This approach is similar in spirit to running BMA/S only on those models which increase their posterior probability relative to the prior. Within this setting, however, we shall not choose a cut-off point based on model prior probability but based on the maximum number of models that we want to include ($topM$) as this is more straightforward. In the next section we shall test different values of $topM$ and show that a handful of pre-screened models is usually enough to have posterior model probability close to 1.

There are two notable ways to handle model pre-screening. First, the SFA model space can be searched faster using only ML estimation, with the set of candidate models chosen based on BIC. Note that as $n \to \text{inf}$, $p(y) \sim -0.5BIC$. Hence, we can run all $2^k$ models using only ML-based estimation and then re-run selected top models ($topM$) using *full* SF-BMA/S algorithm. Pre-screening based on non-Bayesian models and BIC (with Bayesian estimation and Laplace approximation turned off) is much faster and we can use it to filter a set of the most probable models. Unless stated otherwise, $topM$=4000 is set throughout the paper, which is consistent with switching pre-screening whenever $k$>11. The strategy should be especially accurate for larger datasets because the larger $n$, the more BIC should correspond to the real integrated likelihood $p(y)$. For small-to-moderate $n$, and especially for very large $k$, we recommend a larger model set; e.g., $topM$=10000 or even 20000, so as not to miss relevant specifications. The additional computational cost is rather negligible. For example, in one of our empirical examples, where k=27, once we switch to $topM$=20000 the procedure takes just additional five seconds to execute.

Furthermore, as we shall demonstrate in the next section, recovery of the true model is not that much of a problem for simple Gaussian-error BMA/S. What changes are the exact values for model weights, especially in certain scenarios. Thus, we may also construct an even faster pre-screening strategy based on BIC from simple Gaussian-error models, i.e., BMA/S on simple Gaussian-error linear regression models. Since we only need BIC from Gaussian-error models, there is no need for *full* Bayesian estimation of models defined in Eq. (6), and thus we use ML which is available analytically. This way we can perform very fast pre-screening to find a set of potentially relevant models. In our case, using this strategy cuts down the runtime for larger cases ($k > 17$) massively to just a small fraction of what it would otherwise take. It makes the evaluation of models with $k$ equal up to 30 quite feasible. We return to this in the simulations section.

In summary, it should be noted that both model pre-screening strategies are warranted once $k$ reaches high-enough values (e.g. $k$>12). Otherwise, the computational gain is negligible. Also, the rationale behind these pre-screening strategies is that their usefulness is not conditioned upon their accuracy in approximating the true value of integrated likelihood via BIC and thus posterior model probability, nor even on their ability to rank the 'top' models accordingly. It is conditioned only upon their ability to adequately filter a set of the most relevant models with good-enough accuracy (i.e., find models which have a high-enough share of the total posterior model probability). We shall test this in the simulations section.

# 3 Simulation study of the SF-BMA/S performance

## 3.1 Monte Carlo simulation design

For our simulation-based analysis we run 100 Monte Carlo draws with 27 scenarios per draw. These scenarios involve:

- Three cases for the level of parameter $\lambda = \frac{\sigma_u}{\sigma_v}$, which is known in SF literature as the inefficiency(signal)-to-noise ratio (Badunenko and Kumbhakar, 2016; Kumbhakar *et al.*, 2013): $\lambda$={0.5, 1, 4}. Because $\sigma_u$ in SF models has a meaningful interpretation, we fix $\sigma_u$=0.4, which corresponds to mean efficiency in the exponential case of about 0.71, and change $\sigma_v$={0.8, 0.4, 0.1}.
- Three cases for the level of signal strength ratio defined as $SS = \frac{\mathrm{Var}(X\beta^{true})}{\mathrm{Var}(v-u)}$ ={0.25,1,4}, which is the ratio of the variance of the underlying true deterministic signal, also referred to data generating process (DGP: $X\beta^{true}$), and the variance of the compound SF error term ($\varepsilon$).
- Three cases for the number of 'false' regressors: $k_2$={4,8,10}. We consider four 'true' regressors $k_1$=4, which amounts to three scenarios in terms of the total number of candidate variables to choose $k = k_1 + k_2$ ={8,12,14}. The true values of the parameters in the simulation are grouped in a vector of length $k_1$+1, $\beta^{true} = [1; 1; 1.5; 2; -3]$, and the intercept ($\beta_0$) is placed as the first parameter in the vector.

Hence, the study design has 27 scenarios looped over 100 Monte Carlo draws to avoid incidental findings in the simulation results. On top of that, we consider two cases of MC simulation results: one in which the regressors are uncorrelated (the baseline case) and one in which they are moderately correlated (, results are in the appendix). The only constant in the simulation, apart from $\sigma_u$, is the number of observations ($n$), which is set to 500. We choose this number as a compromise between BMA/S and SFA applications. On the one hand, many renown BMA/S applications do not have large $n$; e.g., Fernández et al. (2001) growth dataset available in the BMA package in R has 72 observations, the original dataset by Barro (1991) has 98. On the other hand, however, SFA applications tend to have many more observations, which is obvious as it usually takes more datapoints to accurately measure inefficiency. For example, a well-known production dataset from R's PLM package (Croissant and Millo, 2008), derived from the U.S. state production originally used by Munnell (1990), has $n$=816. The famous WHO dataset on healthcare attainment has $n$=700 (Greene, 2004, 2005, 2018; Hollingsworth and Wildman, 2003; Makieła and Mazur, 2022). The dataset recently used by Makieła et al. (2025) in the context of SFA and BMA/S has $n$=4446.

## 3.2 Accuracy measures selected for Monte Carlo results summary

Based on the abovementioned simulation design, we compare the proposed SF-BMA/S with BMA/S based on standard Gaussian-error linear regression based on Eq. (6), which we shall refer to as Gaussian BMA/S. We evaluate the two BMA/S procedures in three domains: (1) accuracy of true model recovery (i.e., variable/model selection accuracy), (2) accuracy of model averaging, and (3) accuracy of the true parameters' estimation (also referred to as accuracy of the true parameters' recovery).

For accuracy of model recovery (BMS), let $k_1$ be the number of truly relevant regressors and $k_2$ be the number of irrelevant regressors. Further, let $TP$ be the number of true positives (relevant variables correctly selected) and $FP$ the number of false positives (irrelevant variables incorrectly selected). We thus have false negatives as $FN = k_1 - TP$ and true negatives as $TN = k_2 - FP$. This way we can define true-positive rate (fraction of truly relevant regressors recovered) as $TPR = \frac{TP}{k_1}$ and false-positive rate (fraction of irrelevant regressors incorrectly included) as $FPR = \frac{FP}{k_2}$. The indicator for the exact recovery of the true specification per draw (and scenario) is simply $ER = 1$ if the true specification was recovered by the procedure, i.e., if the best model determined by the procedure is the true model given the data-generating process (DGP), 0 otherwise.

For accuracy of model averaging (BMA), we consider average posterior inclusion probabilities (PIP) of true regressors in the model: $PIP_{tr} = \frac{1}{k_1}\sum_{j\in true} PIP_j$, and average posterior inclusion probabilities of false regressors in the model: $PIP_{fs} = \frac{1}{k_2}\sum_{j\in false} PIP_j$, where $PIP_j = \sum_{M:x_j\in M} p(M \mid y)$ is the posterior inclusion probability of variable (regressor) $x_j$. We also consider a measure based on Brier score (Brier, 1950; Gneiting and Raftery, 2007): $BS = \frac{1}{k}\sum_{j=1}^{k}\left(PIP_j - z_j\right)^2$, where $z_j = 1$ if a variable truly belongs to the data-generating process (DGP), 0 otherwise.

In terms of accuracy of the true parameters' recovery, we compare the estimates with the true values using root mean squared error: $RMSE = \sqrt{\frac{1}{k_1}\sum_{j=1}^{k_1}\left(\hat{\beta}_j - \beta_j^{true}\right)^2}$ . Also, to assess recovery of true parameters' signs we count the mean number of cases (per model), in which parameter estimates were of the same sign as the true values: $Sign = \frac{1}{k_1}\sum_{j=1}^{k_1} I\left(\text{sign}\left(\hat{\beta}_j\right) = \text{sign}(\beta_j)\right)$, where $I$ is the indicator function equal 1 if the signs match, 0 otherwise.

All the abovementioned measures are model-specific and need to be computed for every MC draw in every scenario and for both BMA/S procedures. To make the comparison manageable, we report differences between the two procedures averaged over all MC draws for every scenario separately and overall. For readability, the differences are constructed in a way that positive values indicate the advantage of the SF-BMA/S procedure over Gaussian BMA/S, negative values indicate otherwise.

For model recovery comparison we have the mean difference in TPR (true positive rate):

$$\overline{\Delta TPR} = \frac{1}{MC}\sum_{m=1}^{MC}\left(TPR_m^{SF} - TPR_m^{Ga}\right), \tag{7}$$

where $SF$ superscript indicates the stochastic frontier specification and $Ga$ indicates the Gaussian-error case. Positive values indicate that SF-BMA/S has better true positive rates. Next is the negative mean difference in FPR (false positive rate):

$$-\overline{\Delta FPR} = -\frac{1}{MC}\sum_{m=1}^{MC}\left(FPR_m^{SF} - FPR_m^{Ga}\right), \tag{8}$$

and the mean difference in exact model recovery is:

$$\overline{\Delta ER} = \frac{1}{MC}\sum_{m=1}^{MC}\left(ER_m^{SF} - ER_m^{Ga}\right). \tag{9}$$

For model averaging comparison we have the mean difference in average PIP of true regressors:

$$\overline{\Delta PIP}_{tr} = \frac{1}{MC}\sum_{m=1}^{MC}\left({PIP_{tr}}_m^{SF} - {PIP_{tr}}_m^{Ga}\right), \tag{10}$$

the negative mean difference in average posterior inclusion probabilities of false regressors:

$$-\overline{\Delta PIP}_{fs} = -\frac{1}{MC}\sum_{m=1}^{MC}\left({PIP_{fs}}_m^{SF} - {PIP_{fs}}_m^{Ga}\right), \tag{11}$$

and the negative mean difference in Brier score:

$$-\overline{\Delta BS} = -\frac{1}{MC}\sum_{m=1}^{MC}\left(BS_m^{SF} - BS_m^{Ga}\right). \tag{12}$$

Finally, to compare true parameters' estimation accuracy we have the negative mean difference between RMSE's

$$-\overline{\Delta RMSE} = -\frac{1}{MC}\sum_{m=1}^{MC}(RMSE_m^{SF} - RMSE_m^{Ga}), \tag{13}$$

and mean difference in sign recovery

$$\overline{\Delta Sign} = -\frac{1}{MC}\sum_{m=1}^{MC}(Sign_m^{SF} - Sign_m^{Ga}). \tag{14}$$

To sum up, positive mean differences in the accuracy measures reported in the MC simulation summary (Table 1) indicate that, on average, SF-BMA/S performs better, negative values indicate the opposite.

Furthermore, to facilitate a clean, simple comparison for the entire MC simulation, we construct score metrics for every of the abovementioned measures. Since values above zero in a given draw indicate advantage of SF-BMA/S, we define the following score metric for every measure:

$$Score = \frac{1}{MC}\sum_{m=1}^{MC}[I(measure_m > 0) + 0.5\, I(measure_m = 0)], \tag{15}$$

where $I(\cdot)$ denotes the indicator function, so that $I(A) = 1$ if $A$ is true, 0 otherwise. The score interpretation is simple. Values above 0.5 indicate that a given measure was more often in favor of SF-BMA/S, and the closer it is to 1 the more persistent this advantage was throughout the simulation. Values below 0.5 indicate on average advantage of Gaussian BMA/S and values equal to 0.5 indicate a tie (see Table 2).

## 3.3 Monte Carlo simulation results

Results based on the means and the scores for our accuracy measures are summarized in Table 1 and Table 2.

[Table 1 and Table 2 about here]

The SF-BMA/S procedure is almost always better regardless of the accuracy measure. In terms of true model recovery, the results for both procedures are usually close. More notable differences are visible in scenarios with high $\lambda$ ($\lambda$=4), especially when coupled with low signal strength levels (0.25, 1). Whenever inefficiency is well-defined and 'visible' in the data, using the SFA specification is definitely advisable for model recovery. The differences become more substantial when we consider the accuracy of true parameters' estimates or if the task is model averaging. For $\lambda = 4$ some $\overline{\Delta RMSE}$'s are about 0.98 and the score for differences in the average PIP of true regressors is 1 (see scores for $PIP_{tr}$; Table 2), which indicates that in some simulation scenarios PIPs from SF-BMA are always better, with a mean difference of about 14.9% ($PIP_{tr}$; Table 1). Hence, accounting for SFA specification has consequences on model averaging and the accuracy of estimates of the true parameters, especially in some scenarios (e.g. high $\lambda$, low $SS$). The findings remain unchanged regardless of whether correlation between regressors is introduced (see the tables in the Appendix).

On average, Gaussian BMA/S seems to handle model recovery quite well (the overall score is just 0.5459). The question is whether it can be used as a fast pre-screening strategy and how it performs against the *exact* pre-screening strategy based on SF models. To ascertain this, we can also use MC simulation. Let $topM$ be the number of top (best) models we want to use in BMA/S. We can compute (i) what fraction of the $topM$ models ranked based on *proxy* BIC (i.e., BIC from Gaussian-error models) overlaps with the actual $topM$ models ranked based on the highest model posterior probability, and (ii) what fraction of $topM$ models ranked based on *exact* BIC (i.e., BIC from SF models) overlaps with the actual $topM$ models ranked based on the highest model posterior probability. The closer these two values are to one, the more in agreement are the two BIC-based model rankings with the actual one based on posterior model probability. We also report Spearman's rank correlation coefficients between the rankings. More importantly, however, we can sum up the *true* posterior probabilities for the $topM$ models from both BIC rankings based on *proxy* BIC and *exact* BIC. This is the most relevant measure because we do not really care about the order of the rankings (correlation), not even so much about the rate of their overlap. First and foremost, we want to know if the $topM$ models filtered using these two methods 'consume' most of the posterior probability mass, ideally close to one.

Table 3 presents a set of results for the most variable-rich scenarios ($k_2$=10, scenarios: 19-27). These scenarios are the most interesting cases because they provide the largest model pool ($2^{14}$=16384), and thus the biggest potential for error (missing relevant models). The results are grouped based on $topM$={10,100,1000}. Even at $topM$=10 the Gaussian-BIC ranking holds very well, and the results for $topM$=100 are already very close to one. Hence, on condition that $topM$ is not too small, the fast pre-screening strategy is very effective. Unless stated otherwise, we shall use $topM$=4000, which means that fast pre-screening turns on at about $k$>12. This value for $topM$ is too large for our MC simulation but the reader should note that the computational cost of running 4000 SF models is negligible.

[Table 3 about here]

Figure 1 presents computation times (runtimes) of the SF-BMA/S procedure using the fast pre-screening strategy. The procedure was tested on a PC equipped with a 32-core (64-threads) AMD Threadripper 3970X and 64GB of RAM. MATLAB's parpool was set to 50 workers. The computation time is satisfactory up to about $k$=25-27, but still manageable for $k$ up to 30, which corresponds to over 1 billion models searched. The bigger problem is the memory requirement. Fast pre-screening requires us to retain two vectors of length $2^k$ (model IDs and BICs) and a 64GB RAM computer may be out of memory at around $k$=32 (this is depended on other factors like operating system and the size of virtual memory). Hence, we consider $k$=32 as the theoretical maximum for most contemporary PCs, as well as the practical maximum too, because $k$=32 would take over two hours to compute.

[Figure 1 about here]

# 4 Empirical examples

The first empirical example uses the dataset from a recent study by Makieła et al. (2025), which is an unbalanced panel of 4446 observations for annual data on crop production in the EU from 2007 to 2014. The dataset has already been used with BMA/S procedures, but in a different context. Having a pre-selected translog functional form, Makieła et al. (2025) used BMA/S to analyze the underlying distributional assumptions about the error components. Here, we focus on the choice of regressors, which is a more conventional use of BMA/S. The original dataset contains 4 inputs, but we treat the time index as if it was a fifth one, i.e., we add its squared term and allow it to interact with the logs of other inputs in translog, which is analogous to model $m_1$ in Makieła et al. (2025). Hence, we have $k$=20 and $2^k$=1048576 cases (this includes a model with only the intercept).

The results are summarized in Table 4. Since the number of maximum regressors is manageable, we provide two sets or results: (i) based on fast pre-screening and (ii) based on *full* SF-BMA/S (no pre-screening at all). The results are virtually the same with minor differences around the twelfth decimal place. For example, the exact value of |t-ratio| for Time index is 57.1175127398**12** according to SF-BMA/S with fast pre-screening and 57.1175127398**06** based on *full* SF-BMA/S. What actually makes a huge difference is the computation time. SF-BMA/S with fast pre-screening takes about 39 seconds compared to 4371 seconds for *full* SF-BMA/S (and compared to 2494 seconds for SF-BMA/S with *exact* pre-screening based on BIC from SF models).

[Table 4 about here]

The second example is based on the U.S. state production originally used by Munnell (1990). We use the benchmark *Produc* dataset from the PLM package (Croissant and Millo, 2008), which contains $n$=816 observations for the U.S. states over 1970-1986 period and six inputs (i.e., we exclude variable *pcap*, which is the aggregate of public capital inputs: *pcap*=*water*+*hwy*+*util*, and retain its components).

We use the translog production function, which amounts to $k$=27 and 134 217 728 possible model specifications. Since the number of observations is not that big while the number of model combinations is very large, we switch to $topM$=20000 for the fast pre-screening strategy, which takes about 5.5 minutes (322 seconds) to compute. If we were to use $topM$=4000 the computation time would be about 298 seconds, so the additional time needed is rather negligible. The results are summarized in Table 5.

[Table 5 about here]

Given the SF-BMA/S results, we notice that about 40% of the regressors are irrelevant (i.e. 11 out of 27), and this is mostly in the translog extension of the model. Naturally adding or removing regressors may have consequences on inefficiency measurement and that is why SF specifications are generally driven by theory. Hence, low PIPs for the categories of public capital stock like *highway and streets* ($x_3$) or *other public building and structures* ($x_5$) may not necessarily warrant their removal from the model. However, inclusion of squared terms and interaction terms is a technical aspect and should be motivated by model fit. Table 6 shows comparison between efficiency scores from the full translog model and the best model determined by SF-BMA/S (see BMS column in Table 5). The efficiency estimates have been calculated (i) based on classic Jondrow et al. (1982) estimator directly from SF-BMA/S results as well as (ii) outside the procedure using Gibbs sampler (Makieła, 2017, 2020).

[Table 6 about here]

Although the relationship between the two efficiency rankings is almost perfectly monotonic (Spearman's rank correlation is about 0.98), Kendall's tau is about 0.88, which indicated that there is a noticeable number of pairwise rank reversals. Moreover, although the best model has 11 regressors fewer to explain the variation of the dependent compared to the full model, its average technical efficiency score in the sample is still marginally higher. This is consistent with the findings in Makieła and Osiewalski (2018), who notice that the best SF models given the data show higher average efficiency. We reaffirm their finding that the choice of the optimal parametrization of the production or cost function, especially with respect to regressor choice, may have nonnegligible consequences for efficiency measurement. This therefore motivates the use of the model search methods, such as the one proposed in this paper.

Finally, we find that the mean efficiency is relatively high under both rankings, which may raise concerns that the marginal posterior distributions of the efficiency scores are concentrated near one, which indicates full efficiency. Figure 1 shows these posteriors in the best model for three representative cases: the least efficient observation, the most efficient observation, and the observation closest to the mean efficiency. In two out of three cases, the posterior mode is visibly below one, indicating that the posterior mass is not concentrated at full efficiency. This provides support for the presence of inefficiency in the analyzed data.

[Figure 2 about here]

# 5 Concluding remarks

The paper focuses on covariate selection in Stochastic Frontier Analysis (SFA). This issue is particularly important in SFA because efficiency estimates are conditional on the parametric specification of the frontier function. In this aspect, the study complements the work of Makieła and Mazur (2022), who employed Bayesian Model Averaging in SFA to address uncertainty regarding the specification of the compound error term rather than uncertainty associated with covariate selection. More broadly, the paper extends the research initiated by Makieła and Osiewalski (2018), who were among the first to apply Bayesian Model Averaging to the covariate selection problem in SFA.

In SFA, model specification is typically guided by economic theory, which means that the covariate selection problem is often considerably smaller compared to settings where all available variables are treated as potential candidates. For example, in case of translog functional forms, BMA/S in SFA would likely be used primarily on the elements of the second-order approximation, which substantially reduces the total number of possible model combinations. This motivates us to use exhaustive search algorithms with parallelization, which may be more convenient when the number of candidates ($k$) is not too large. The proposed framework has been tested for $k$ of up to 30 (i.e., slightly over one billion models), which takes about 31.5 minutes to compute. For a computer equipped with 64GB of RAM, the theoretically upper limit for the proposed strategy with exhaustive search and fast pre-screening is around $k$=32 and it would take slightly over two hours of runtime. Consequently, most contemporary computers should be able to handle cases of $k$ up to 29. It is worth noting that the hardware used in the study (32-core AMD Threadripper 3970X, 64GB RAM), although still potent, dates back to 2020.

Monte Carlo results show that relying on SF specification over Gaussian-error specification is recommended for BMA as well as for parameter recovery whenever inefficiency-to-nose ratio is high ($\lambda > 1$) and the underlying signal strength of the data generating process is not large ($SS \leq 1$). Arguably, however, these are also the scenarios in which efficiency evaluation is most relevant. With respect to model selection (BMS), both procedures perform comparably. That is, the advantage of SF-BMA/S is rather consistent across the scenarios analyzed but usually negligible.

Further Monte Carlo analysis indicates that it is justified to use a fast pre-screening strategy based on classical Gaussian-error models and their BICs. This drastically improves computational speed without having much impact on the coverage of relevant models. It is also worth noting that SFA models can become numerically unstable in terms of nonlinear likelihood which can have almost flat regions. This makes estimation difficult in some extreme cases, particularly for poorly specified regressor sets. Hence, restricting the analysis to a preselected subset of top-performing models, obtained with a more numerically stable procedure, makes even more sense.

Although efficiency analysis is the focal point in SFA, efficiency estimation itself is not the primary concern of this paper. First, Gaussian BMA/S does not permit inefficiency, which makes the disadvantage compared to SF-BMA/S self-evident. Second, obtaining the (in)efficiency scores once we have information from SF-BMA/S is rather straightforward and depends on the strategy one wants to pursue. If the goal is efficiency analysis based on a single best model specification (BMS), a good option would be to follow one of our empirical examples and, e.g., use an external package for 'full' estimation of the selected model (Makieła, 2017, 2020). If the goal is BMA, one can use Jondrow et al. (1982) estimator, available in SF-BMA/S, and compute efficiency scores in a *parfor* loop over a set of relevant models. In practice, this subset would likely be restricted to approximately 100–1000 models because of memory constraints associated with storing large numbers of efficiency estimates.

Finally, the paper focuses on normal-exponential SF models. Nevertheless, it is worth mentioning that the proposed SF-BMA/S framework can also handle normal-half-normal and panel data SFA cases as well as Bayesian Random Effects models. Although panel data modelling is outside the scope of this study, accounting for panel data structure would likely strengthen the results in favor of SF-BMA/S. We leave this issue for future research.

## Code Availability

The methods discussed in this paper are implemented in both MATLAB and Python. The MATLAB package includes all procedures considered in the paper, as well as several additional functionalities. The Python package currently supports estimation of individual SF models.

The **MATLAB** package is available at: https://github.com/KamilMakiela/SF-BMA
The **Python** package is available at: https://github.com/KamilMakiela/Python-SFA

# References

[1] Badunenko, O. and Kumbhakar, S.C. (2016), “When, where and how to estimate persistent and transient efficiency in stochastic frontier panel data models”, *European Journal of Operational Research*, Vol. 255 No. 1, pp. 272–287, doi: 10.1016/j.ejor.2016.04.049.

[2] Barro, R.J. (1991), “Economic Growth in a Cross Section of Countries”, *The Quarterly Journal of Economics*, Vol. 106 No. 2, p. 407, doi: 10.2307/2937943.

[3] Battese, G.E. and Corra, G.S. (1977), “Estimation of a Production Frontier Model: With Application to the Pastoral Zone of Eastern Australia”, *Australian Journal of Agricultural Economics*, Vol. 21 No. 3, pp. 169–179, doi: 10.1111/j.1467-8489.1977.tb00204.x.

[4] Brier, G.W. (1950), “Verification of Forecasts Expressed in Terms of Probability”, *Monthly Weather Review*, Vol. 78 No. 1, pp. 1–3, doi: 10.1175/1520-0493(1950)078<0001:VOFEIT>2.0.CO;2.

[5] van den Broeck, J., Koop, G., Osiewalski, J. and Steel, M.F.J. (1994), “Stochastic frontier models: a Bayesian perspective”, *Journal of Econometrics*, Vol. 61 No. 2, pp. 273–303, doi: 10.1016/0304-4076(94)90087-6.

[6] Croissant, Y. and Millo, G. (2008), “Panel Data Econometrics in *R* : The **plm** Package”, *Journal of Statistical Software*, Vol. 27 No. 2, doi: 10.18637/jss.v027.i02.

[7] Fernández, C., Ley, E. and Steel, M.F.J. (2001), “Benchmark priors for Bayesian model averaging”, *Journal of Econometrics*, Vol. 100 No. 2, pp. 381–427, doi: 10.1016/S0304-4076(00)00076-2.

[8] George, E.I. and McCulloch, R.E. (1993), “Variable Selection via Gibbs Sampling”, *Journal of the American Statistical Association*, Vol. 88 No. 423, pp. 881–889, doi: 10.1080/01621459.1993.10476353.

[9] Gneiting, T. and Raftery, A.E. (2007), “Strictly Proper Scoring Rules, Prediction, and Estimation”, *Journal of the American Statistical Association*, Vol. 102 No. 477, pp. 359–378, doi: 10.1198/016214506000001437.

[10] Greene, W. (2004), “Distinguishing between heterogeneity and inefficiency: stochastic frontier analysis of the World Health Organization’s panel data on national health care systems”, *Health Economics*, Vol. 13 No. 10, pp. 959–980, doi: 10.1002/hec.938.

[11] Greene, W. (2005), “Fixed and Random Effects in Stochastic Frontier Models”, *Journal of Productivity Analysis*, Vol. 23 No. 1, pp. 7–32, doi: 10.1007/s11123-004-8545-1.

[12] Greene, W. (2018), *Econometric Analysis*, 8th ed., Prentice Hall / Pearson, Upper Saddle River, NJ.

[13] Hoeting, J.A., Madigan, D., Raftery, A.E. and Volinsky, C.T. (1999), “Bayesian model averaging: a tutorial (with comments by M. Clyde, David Draper and E. I. George, and a rejoinder by the authors”, *Statistical Science*, Vol. 14 No. 4, doi: 10.1214/ss/1009212519.

[14] Hollingsworth, B. and Wildman, J. (2003), “The efficiency of health production: re-estimating the WHO panel data using parametric and non-parametric approaches to provide additional information”, *Health Economics*, Vol. 12 No. 6, pp. 493–504, doi: 10.1002/hec.751.

[15] Jondrow, J., Knox Lovell, C.A., Materov, I.S. and Schmidt, P. (1982), "On the estimation of technical inefficiency in the stochastic frontier production function model", *Journal of Econometrics*, Vol. 19 No. 2–3, pp. 233–238, doi: 10.1016/0304-4076(82)90004-5.

[16] Kumbhakar, S.C., Parmeter, C.F. and Tsionas, E.G. (2013), "A zero inefficiency stochastic frontier model", *Journal of Econometrics*, Vol. 172 No. 1, pp. 66–76, doi: 10.1016/j.jeconom.2012.08.021.

[17] Ley, E. and Steel, M.F.J. (2009), "On the effect of prior assumptions in Bayesian model averaging with applications to growth regression", *Journal of Applied Econometrics*, Vol. 24 No. 4, pp. 651–674, doi: 10.1002/jae.1057.

[18] Li, G., Liu, Z., Zhang, J., Han, H. and Shu, Z. (2024), "Bayesian model averaging by combining deep learning models to improve lake water level prediction", *Science of The Total Environment*, Vol. 906, p. 167718, doi: 10.1016/j.scitotenv.2023.167718.

[19] Makieła, K. (2014), "Bayesian Stochastic Frontier Analysis of Economic Growth and Productivity Change in the EU, USA, Japan and Switzerland", *Central European Journal of Economic Modelling and Econometrics*, Vol. 6 No. 3, pp. 193–216, doi: https://doi.org/10.24425/cejeme.2014.119239.

[20] Makieła, K. (2017), "Bayesian Inference and Gibbs Sampling in Generalized True Random-Effects Models", *Central European Journal of Economic Modelling and Econometrics*, Vol. 9 No. 1, pp. 69–95, doi: https://doi.org/10.24425/cejeme.2017.122200.

[21] Makieła, K. (2020), "MATLAB Function for Bayesian Stochastic Frontier Analysis".

[22] Makieła, K., Marzec, J., Pisulewski, A. and Mazur, B. (2025), "Are European Farms Equally Efficient? What Do Regional FADN Data on Crop Farms Tell Us?", *Journal of Agricultural and Applied Economics*, Vol. 57 No. 1, pp. 157–181, doi: 10.1017/aae.2025.1.

[23] Makieła, K. and Mazur, B. (2020), "Bayesian Model Averaging and Prior Sensitivity in Stochastic Frontier Analysis", *Econometrics*, Vol. 8 No. 2, p. 13, doi: 10.3390/econometrics8020013.

[24] Makieła, K. and Mazur, B. (2022), "Model uncertainty and efficiency measurement in stochastic frontier analysis with generalized errors", *Journal of Productivity Analysis*, Vol. 58 No. 1, pp. 35–54, doi: 10.1007/s11123-022-00639-y.

[25] Makieła, K. and Osiewalski, J. (2018), "Cost Efficiency Analysis of Electricity Distribution Sector under Model Uncertainty", *The Energy Journal*, Vol. 39 No. 4, pp. 31–56, doi: 10.5547/01956574.39.4.kmak.

[26] Miller, A. (2002), *Subset Selection in Regression*, Chapman and Hall/CRC, doi: 10.1201/9781420035933.

[27] Mitchell, T.J. and Beauchamp, J.J. (1988), "Bayesian Variable Selection in Linear Regression", *Journal of the American Statistical Association*, Vol. 83 No. 404, pp. 1023–1032, doi: 10.1080/01621459.1988.10478694.

[28] Munnell, A.H. (1990), "Why Has Productivity Growth Declined? Productivity and Public Investment", *New England Economic Review*, pp. 3–22.

[29] Osiewalski, J. and Steel, M.F.J. (1993), "Regression Models under Competing Covariance Structures: A Bayesian Perspective", *Annales d'Economie et de Statistique*, No. 32, pp. 65–79.

[30] Parmeter, C.F. and Kumbhakar, S.C. (2014), “Efficiency Analysis: A Primer on Recent Advances”, *Foundations and Trends® in Econometrics*, Vol. 7 No. 3–4, pp. 191–385, doi: 10.1561/0800000023.

[31] Raftery, A.E., Madigan, D. and Hoeting, J.A. (1997), “Bayesian Model Averaging for Linear Regression Models”, *Journal of the American Statistical Association*, Vol. 92 No. 437, pp. 179–191, doi: 10.1080/01621459.1997.10473615.

[32] Raftery, A.E., Painter, I.S. and Volinsky, C.T. (2005), “BMA: An R Package for Bayesian Model Averaging”, *R News*, Vol. 5 No. 2, pp. 2–8.

[33] Smith, M. and Kohn, R. (1996), “Nonparametric regression using Bayesian variable selection”, *Journal of Econometrics*, Vol. 75 No. 2, pp. 317–343, doi: 10.1016/0304-4076(95)01763-1.

[34] Steel, M.F.J. (2020), “Model Averaging and Its Use in Economics”, *Journal of Economic Literature*, Vol. 58 No. 3, pp. 644–719, doi: 10.1257/jel.20191385.

[35] Steiner, G. and Steel, M. (2026), “Bayesian Model Averaging in Causal Instrumental Variable Models”, *Journal of Applied Econometrics*, doi: 10.1002/jae.70070.

[36] Tasiopoulos, A.E., Tsionas, E.G. and Vlastakis, N.D. (2026), “Bayesian model averaging with non-conjugate priors”, *Journal of Econometrics*, p. 106256, doi: 10.1016/j.jeconom.2026.106256.

[37] Tsionas, E.G. (2012), “Maximum likelihood estimation of stochastic frontier models by the Fourier transform”, *Journal of Econometrics*, Vol. 170 No. 1, pp. 234–248, doi: 10.1016/j.jeconom.2012.04.001.

[38] Zellner, A. (1971). *An Introduction to Bayesian Inference in Econometrics*. New York: John Wiley & Sons.

# Tables and Figures

**Table 1: Mean differences between accuracy measures in SF-BMA/S and Gaussian BMA/S**

| Sc. | $k_2$ | $\lambda$ | SS | $\overline{\Delta TPR}$ | $-\overline{\Delta FPR}$ | $\Delta ER$ | $\overline{\Delta PIP}_{tr}$ | $-\overline{\Delta PIP}_{fs}$ | $-\overline{\Delta BS}$ | $-\overline{\Delta RMSE}$ | $\overline{\Delta Sign}$ |
|---|---|---|---|---|---|---|---|---|---|---|---|
| 1 | 4 | 0.5 | 0.25 | 0 | 0 | 0.01 | 0.0013 | -0.0006 | 0.0003 | 0.0016 | 0 |
| 2 | 4 | 0.5 | 1 | -0.005 | 0 | -0.02 | 0.0013 | 0.0001 | 0.0000 | 0.0005 | 0 |
| 3 | 4 | 0.5 | 4 | 0 | 0 | 0 | 0.0000 | -0.0003 | -0.0001 | 0.0001 | 0 |
| 4 | 4 | 1 | 0.25 | 0.013 | 0.005 | 0.01 | 0.0153 | 0.0054 | 0.0085 | 0.0154 | 0.0025 |
| 5 | 4 | 1 | 1 | 0.0225 | -0.0025 | 0.08 | 0.0155 | 0.0020 | 0.0056 | 0.0104 | 0 |
| 6 | 4 | 1 | 4 | 0 | 0.0025 | 0.01 | 0.0000 | 0.0017 | 0.0000 | 0.0029 | 0 |
| 7 | 4 | 4 | 0.25 | 0.155 | 0.0025 | 0.53 | 0.1384 | 0.0207 | 0.0622 | 0.1124 | 0.0075 |
| 8 | 4 | 4 | 1 | 0.035 | 0.0025 | 0.15 | 0.0375 | 0.0107 | 0.0119 | 0.0447 | 0 |
| 9 | 4 | 4 | 4 | 0 | 0.0025 | 0.01 | 0.0000 | 0.0082 | 0.0015 | 0.0201 | 0 |
| 10 | 8 | 0.5 | 0.25 | 0.0075 | -0.003 | 0.02 | 0.0014 | 0.0001 | 0.0002 | 0.0020 | 0 |
| 11 | 8 | 0.5 | 1 | -0.0025 | 0 | -0.01 | 0.0006 | -0.0003 | -0.0001 | 0.0003 | 0 |
| 12 | 8 | 0.5 | 4 | 0 | 0 | 0 | 0.0000 | 0.0001 | 0.0000 | 0.0002 | 0 |
| 13 | 8 | 1 | 0.25 | 0.008 | -0.0013 | 0 | 0.0210 | 0.0039 | 0.0040 | 0.0156 | 0.005 |
| 14 | 8 | 1 | 1 | 0.0125 | 0 | 0.05 | 0.0123 | 0.0019 | 0.0029 | 0.0078 | 0 |
| 15 | 8 | 1 | 4 | 0 | 0 | 0 | 0.0000 | 0.0018 | 0.0003 | 0.0023 | 0 |
| 16 | 8 | 4 | 0.25 | 0.175 | 0.0113 | 0.64 | 0.1517 | 0.0234 | 0.0468 | 0.0980 | 0 |
| 17 | 8 | 4 | 1 | 0.035 | 0.006 | 0.17 | 0.0423 | 0.0133 | 0.0110 | 0.0434 | 0 |
| 18 | 8 | 4 | 4 | 0 | 0.0025 | 0.02 | 0.0001 | 0.0050 | 0.0007 | 0.0131 | 0 |
| 19 | 10 | 0.5 | 0.25 | 0.0025 | 0 | 0.01 | 0.0008 | 0.0014 | 0.0011 | 0.0012 | 0 |
| 20 | 10 | 0.5 | 1 | 0 | 0 | 0 | 0.0010 | 0.0004 | 0.0004 | 0.0006 | 0 |
| 21 | 10 | 0.5 | 4 | 0 | 0 | 0 | 0.0000 | 0.0000 | -0.0001 | 0.0000 | 0 |
| 22 | 10 | 1 | 0.25 | 0.023 | 0.001 | 0.03 | 0.0202 | 0.0028 | 0.0049 | 0.0126 | 0.005 |
| 23 | 10 | 1 | 1 | 0.0025 | -0.001 | 0 | 0.0125 | 0.0021 | 0.0024 | 0.0054 | 0 |
| 24 | 10 | 1 | 4 | 0 | 0 | 0 | 0.0000 | 0.0000 | 0.0000 | 0.0028 | 0 |
| 25 | 10 | 4 | 0.25 | 0.165 | 0.004 | 0.6 | 0.1532 | 0.0210 | 0.0392 | 0.0905 | 0.0075 |
| 26 | 10 | 4 | 1 | 0.04 | 0.001 | 0.16 | 0.0491 | 0.0083 | 0.0085 | 0.0426 | 0 |
| 27 | 10 | 4 | 4 | 0 | 0.004 | 0.04 | 0.0000 | 0.0086 | 0.0014 | 0.0138 | 0 |
| | | | **Overall** | **0.0255** | **0.0014** | **0.0930** | **0.0250** | **0.0052** | **0.0079** | **0.0207** | **0.0010** |

*Notes*: Label ‘Sc.’ indicates the scenario number.

**Table 2: Scores between accuracy measures in SF-BMA/S and Gaussian BMA/S**

| Sc. | $k_2$ | $\lambda$ | SS | $\overline{\Delta TPR}$ | $-\overline{\Delta FPR}$ | $\Delta ER$ | $\overline{\Delta PIP}_{tr}$ | $-\overline{\Delta PIP}_{fs}$ | $-\overline{\Delta BS}$ | $-\overline{\Delta RMSE}$ | $\overline{\Delta Sign}$ |
|---|---|---|---|---|---|---|---|---|---|---|---|
| 1 | 4 | 0.5 | 0.25 | 0.50 | 0.50 | 0.51 | 0.46 | 0.55 | 0.43 | 0.57 | 0.50 |
| 2 | 4 | 0.5 | 1 | 0.49 | 0.50 | 0.49 | 0.56 | 0.57 | 0.56 | 0.54 | 0.50 |
| 3 | 4 | 0.5 | 4 | 0.50 | 0.50 | 0.50 | 0.56 | 0.59 | 0.59 | 0.48 | 0.50 |
| 4 | 4 | 1 | 0.25 | 0.52 | 0.51 | 0.51 | 0.63 | 0.66 | 0.65 | 0.68 | 0.51 |
| 5 | 4 | 1 | 1 | 0.55 | 0.50 | 0.54 | 0.78 | 0.64 | 0.70 | 0.66 | 0.50 |
| 6 | 4 | 1 | 4 | 0.50 | 0.51 | 0.51 | 0.86 | 0.66 | 0.65 | 0.69 | 0.50 |
| 7 | 4 | 4 | 0.25 | 0.80 | 0.51 | 0.77 | 0.93 | 0.86 | 0.89 | 0.92 | 0.52 |
| 8 | 4 | 4 | 1 | 0.57 | 0.51 | 0.58 | 0.99 | 0.87 | 0.88 | 0.88 | 0.50 |
| 9 | 4 | 4 | 4 | 0.50 | 0.51 | 0.51 | 0.95 | 0.88 | 0.83 | 0.89 | 0.50 |
| 10 | 8 | 0.5 | 0.25 | 0.52 | 0.49 | 0.51 | 0.53 | 0.57 | 0.54 | 0.57 | 0.50 |
| 11 | 8 | 0.5 | 1 | 0.50 | 0.50 | 0.50 | 0.52 | 0.54 | 0.55 | 0.57 | 0.50 |
| 12 | 8 | 0.5 | 4 | 0.50 | 0.50 | 0.50 | 0.59 | 0.63 | 0.60 | 0.56 | 0.50 |
| 13 | 8 | 1 | 0.25 | 0.52 | 0.50 | 0.50 | 0.66 | 0.68 | 0.65 | 0.62 | 0.51 |
| 14 | 8 | 1 | 1 | 0.53 | 0.50 | 0.53 | 0.74 | 0.63 | 0.67 | 0.67 | 0.50 |
| 15 | 8 | 1 | 4 | 0.50 | 0.50 | 0.50 | 0.83 | 0.64 | 0.63 | 0.63 | 0.50 |
| 16 | 8 | 4 | 0.25 | 0.83 | 0.54 | 0.82 | 0.97 | 0.83 | 0.97 | 0.97 | 0.50 |
| 17 | 8 | 4 | 1 | 0.57 | 0.52 | 0.59 | 1.00 | 0.84 | 0.83 | 0.95 | 0.50 |
| 18 | 8 | 4 | 4 | 0.50 | 0.51 | 0.51 | 0.93 | 0.79 | 0.74 | 0.91 | 0.50 |
| 19 | 10 | 0.5 | 0.25 | 0.51 | 0.50 | 0.51 | 0.48 | 0.59 | 0.54 | 0.51 | 0.50 |
| 20 | 10 | 0.5 | 1 | 0.50 | 0.50 | 0.50 | 0.57 | 0.53 | 0.54 | 0.54 | 0.50 |
| 21 | 10 | 0.5 | 4 | 0.50 | 0.50 | 0.50 | 0.64 | 0.54 | 0.55 | 0.54 | 0.50 |
| 22 | 10 | 1 | 0.25 | 0.55 | 0.51 | 0.52 | 0.72 | 0.58 | 0.65 | 0.56 | 0.51 |
| 23 | 10 | 1 | 1 | 0.51 | 0.50 | 0.50 | 0.79 | 0.71 | 0.72 | 0.66 | 0.50 |
| 24 | 10 | 1 | 4 | 0.50 | 0.50 | 0.50 | 0.84 | 0.61 | 0.56 | 0.60 | 0.50 |
| 25 | 10 | 4 | 0.25 | 0.80 | 0.52 | 0.80 | 0.96 | 0.84 | 0.94 | 0.95 | 0.52 |
| 26 | 10 | 4 | 1 | 0.58 | 0.51 | 0.58 | 1.00 | 0.85 | 0.85 | 0.92 | 0.50 |
| 27 | 10 | 4 | 4 | 0.50 | 0.52 | 0.52 | 0.86 | 0.88 | 0.81 | 0.89 | 0.50 |
| | | | **Overall** | **0.548** | **0.505** | **0.546** | **0.753** | **0.687** | **0.686** | **0.701** | **0.502** |

*Notes*: Label 'Sc.' indicates the scenario number.

**Table 3: Comparison of top model rankings based on BIC from SF models and BIC from Gaussian-error models with models' actual posterior probabilities**

| | | | | Total posterior probability of $topM$ models | | Fraction of models that overlap with the true ranking | | Spearman's rank correlation with the true model ranking | |
|---|---|---|---|---|---|---|---|---|---|
| Sc. | $k_2$ | $\lambda$ | SS | SFA | Gaussian | SFA | Gaussian | SFA | Gaussian |
| **$topM$=10** | | | | | | | | | |
| 19 | 10 | 0.5 | 0.25 | 0.902 | 0.743 | 0.805 | 0.410 | 0.987 | 0.776 |
| 20 | 10 | 0.5 | 1 | 0.969 | 0.943 | 0.843 | 0.750 | 0.991 | 0.971 |
| 21 | 10 | 0.5 | 4 | 0.984 | 0.961 | 0.791 | 0.755 | 0.998 | 0.997 |
| 22 | 10 | 1 | 0.25 | 0.821 | 0.799 | 0.853 | 0.753 | 0.972 | 0.965 |
| 23 | 10 | 1 | 1 | 0.915 | 0.914 | 0.799 | 0.778 | 0.988 | 0.989 |
| 24 | 10 | 1 | 4 | 0.976 | 0.976 | 0.858 | 0.872 | 0.996 | 0.996 |
| 25 | 10 | 4 | 0.25 | 0.799 | 0.795 | 0.836 | 0.834 | 0.975 | 0.975 |
| 26 | 10 | 4 | 1 | 0.924 | 0.921 | 0.848 | 0.836 | 0.989 | 0.989 |
| 27 | 10 | 4 | 4 | 0.977 | 0.979 | 0.857 | 0.869 | 0.995 | 0.995 |
| **$topM$=100** | | | | | | | | | |
| 19 | 10 | 0.5 | 0.25 | 0.996 | 0.980 | 0.865 | 0.555 | 0.987 | 0.776 |
| 20 | 10 | 0.5 | 1 | 1.000 | 0.998 | 0.943 | 0.685 | 0.991 | 0.971 |
| 21 | 10 | 0.5 | 4 | 1.000 | 1.000 | 0.933 | 0.742 | 0.998 | 0.997 |
| 22 | 10 | 1 | 0.25 | 0.985 | 0.982 | 0.888 | 0.841 | 0.972 | 0.965 |
| 23 | 10 | 1 | 1 | 0.997 | 0.997 | 0.866 | 0.845 | 0.988 | 0.989 |
| 24 | 10 | 1 | 4 | 1.000 | 1.000 | 0.954 | 0.873 | 0.996 | 0.996 |
| 25 | 10 | 4 | 0.25 | 0.982 | 0.981 | 0.901 | 0.890 | 0.975 | 0.975 |
| 26 | 10 | 4 | 1 | 0.998 | 0.998 | 0.865 | 0.861 | 0.989 | 0.989 |
| 27 | 10 | 4 | 4 | 1.000 | 1.000 | 0.962 | 0.952 | 0.995 | 0.995 |
| **$topM$=1000** | | | | | | | | | |
| 19 | 10 | 0.5 | 0.25 | 1.000 | 1.000 | 0.896 | 0.714 | 0.987 | 0.776 |
| 20 | 10 | 0.5 | 1 | 1.000 | 1.000 | 0.913 | 0.815 | 0.991 | 0.971 |
| 21 | 10 | 0.5 | 4 | 1.000 | 1.000 | 0.998 | 0.991 | 0.998 | 0.997 |
| 22 | 10 | 1 | 0.25 | 1.000 | 1.000 | 0.915 | 0.901 | 0.972 | 0.965 |
| 23 | 10 | 1 | 1 | 1.000 | 1.000 | 0.880 | 0.907 | 0.988 | 0.989 |
| 24 | 10 | 1 | 4 | 1.000 | 1.000 | 0.975 | 0.976 | 0.996 | 0.996 |
| 25 | 10 | 4 | 0.25 | 1.000 | 1.000 | 0.925 | 0.924 | 0.975 | 0.975 |
| 26 | 10 | 4 | 1 | 1.000 | 1.000 | 0.897 | 0.901 | 0.989 | 0.989 |
| 27 | 10 | 4 | 4 | 1.000 | 1.000 | 0.953 | 0.953 | 0.995 | 0.995 |

*Notes*: Averages based on 50 MC draws; label 'Sc.' indicates the scenario number; Spearman's rank correlation was calculated for models which are in both rankings (i.e., the models that overlap in both rankings: the BIC-based one and the *exact,* posterior probability-based, one).

**Table 4: SF-BMA/S results: crop farming in the EU**

| | Fast pre-screening with $topM$=400 (execution time: 39.2 seconds) | | | | | Exact SF-BMA, no pre-screening (execution time: 4371.3 seconds) | | | | | |
|---|---|---|---|---|---|---|---|---|---|---|---|
| ***var.*** | **PIP** | **Est.** | **SE** | **\|*t*-ratio\|** | **PSP** | **PIP** | **Est.** | **SE** | **\|*t*-ratio\|** | **PSP** | **BMS** |
| const | 1 | 11.154 | 0.005 | 2461.57 | 1 | 1 | 11.154 | 0.005 | 2461.57 | 1 | 1 |
| $x_1$ | 1 | 0.092 | 0.003 | 27.258 | 1 | 1 | 0.092 | 0.003 | 27.258 | 1 | 1 |
| $x_2$ | 1 | 0.236 | 0.007 | 35.553 | 1 | 1 | 0.236 | 0.007 | 35.553 | 1 | 1 |
| $x_3$ | 1 | 0.838 | 0.006 | 144.300 | 1 | 1 | 0.838 | 0.006 | 144.300 | 1 | 1 |
| $x_4$ | 1 | -0.142 | 0.006 | 22.848 | 0 | 1 | -0.142 | 0.006 | 22.848 | 0 | 1 |
| $x_5$ | 1 | -0.058 | 0.001 | 57.118 | 0 | 1 | -0.058 | 0.001 | 57.118 | 0 | 1 |
| $x_1^2$ | 1 | 0.091 | 0.002 | 39.706 | 1 | 1 | 0.091 | 0.002 | 39.706 | 1 | 1 |
| $x_2^2$ | 1 | -0.069 | 0.004 | 16.315 | 0 | 1 | -0.069 | 0.004 | 16.315 | 0 | 1 |
| $x_3^2$ | 0.987 | 0.055 | 0.008 | 7.281 | 0.994 | 0.987 | 0.055 | 0.008 | 7.281 | 0.994 | 1 |
| $x_4^2$ | 0.974 | 0.038 | 0.007 | 5.572 | 0.987 | 0.974 | 0.038 | 0.007 | 5.572 | 0.987 | 1 |
| $x_5^2$ | 1 | 0.003 | 0.000 | 10.686 | 1 | 1 | 0.003 | 0.000 | 10.686 | 1 | 1 |
| $x_1x_2$ | 0.988 | -0.060 | 0.010 | 5.706 | 0.006 | 0.988 | -0.060 | 0.010 | 5.706 | 0.006 | 1 |
| $x_1x_3$ | 1 | -0.104 | 0.010 | 10.094 | 0 | 1 | -0.104 | 0.010 | 10.094 | 0 | 1 |
| $x_1x_4$ | 0.003 | 0.000 | 0.001 | 0.030 | 0.501 | 0.003 | 0.000 | 0.001 | 0.030 | 0.501 | 0 |
| $x_1x_5$ | 0.001 | 0.000 | 0.000 | 0.030 | 0.501 | 0.001 | 0.000 | 0.000 | 0.030 | 0.501 | 0 |
| $x_2x_3$ | 1 | 0.126 | 0.013 | 9.423 | 1 | 1.000 | 0.126 | 0.013 | 9.423 | 1 | 1 |
| $x_2x_4$ | 0.001 | 0.000 | 0.001 | 0.022 | 0.5 | 0.001 | 0.000 | 0.001 | 0.022 | 0.5 | 0 |
| $x_2x_5$ | 1 | -0.011 | 0.001 | 7.746 | 0 | 1 | -0.011 | 0.001 | 7.746 | 0 | 1 |
| $x_3x_4$ | 1 | -0.098 | 0.014 | 7.038 | 0 | 1 | -0.098 | 0.014 | 7.038 | 0 | 1 |
| $x_3x_5$ | 0 | 0.000 | 0.000 | 0.012 | 0.5 | 0 | 0.000 | 0.000 | 0.012 | 0.5 | 0 |
| $x_4x_5$ | 1 | 0.009 | 0.001 | 8.321 | 1 | 1 | 0.009 | 0.001 | 8.321 | 1 | 1 |
| $\sigma_v$ | 1 | 0.255 | 0.002 | 104.362 | 1 | 1 | 0.255 | 0.002 | 104.362 | 1 | 1 |
| $\sigma_u$ | 1 | 0.044 | 0.016 | 2.705 | 1 | 1 | 0.044 | 0.016 | 2.705 | 1 | 1 |

*Notes*: PSP is Posterior Sign Probability; BMS column indicates covariates selected for the best model based on the highest model posterior probability; $x_1$ is buildings and machinery; $x_2$ is labor; $x_3$ is materials; $x_4$ is land; $x_5$ is time index; the dependent is crop production; see Makieła et al. (2025) for details; all variables are in natural logs and have been mean-corrected so that first order approximations represent elasticities at the sample mean (i.e., geometric mean). Point estimates of BMA are averaged based on posterior modes.

**Table 5: SF-BMA/S results: the US state production**

| *var.* | PIP | Est | SE | \|*t*-ratio\| | PSP | BMS |
|---|---|---|---|---|---|---|
| const | 1 | 10.509 | 0.004 | 2936.9 | 1.00 | 1 |
| $x_1$ | 1 | 0.651 | 0.005 | 137.0 | 1.00 | 1 |
| $x_2$ | 1 | 0.268 | 0.007 | 40.65 | 1.00 | 1 |
| $x_3$ | 0.001 | 0.000 | 0.000 | 0.014 | 0.50 | 0 |
| $x_4$ | 1 | 0.121 | 0.010 | 12.17 | 1.00 | 1 |
| $x_5$ | 0.002 | 0.000 | 0.001 | 0.003 | 0.50 | 0 |
| $x_6$ | 1 | -0.042 | 0.007 | 5.833 | 0.00 | 1 |
| $x_1^2$ | 0.998 | 0.181 | 0.023 | 8.044 | 1.00 | 1 |
| $x_2^2$ | 1 | 0.251 | 0.029 | 8.610 | 1.00 | 1 |
| $x_3^2$ | 0 | 0.001 | 0.011 | 0.105 | 0.51 | 0 |
| $x_4^2$ | 0 | 0.000 | 0.001 | 0.040 | 0.50 | 0 |
| $x_5^2$ | 0.109 | 0.005 | 0.014 | 0.329 | 0.55 | 0 |
| $x_6^2$ | 0 | -0.002 | 0.010 | 0.203 | 0.48 | 0 |
| $x_1x_2$ | 1 | -0.692 | 0.080 | 8.608 | 0.00 | 1 |
| $x_1x_3$ | 0.932 | 0.352 | 0.122 | 2.894 | 0.97 | 1 |
| $x_1x_4$ | 0.005 | 0.000 | 0.005 | 0.039 | 0.50 | 0 |
| $x_1x_5$ | 0.139 | -0.005 | 0.017 | 0.286 | 0.44 | 0 |
| $x_1x_6$ | 0 | -0.002 | 0.013 | 0.158 | 0.49 | 0 |
| $x_2x_3$ | 0.641 | -0.096 | 0.079 | 1.203 | 0.18 | 1 |
| $x_2x_4$ | 1 | 0.280 | 0.039 | 7.103 | 1.00 | 1 |
| $x_2x_5$ | 1 | -0.052 | 0.057 | 0.901 | 0.14 | 1 |
| $x_2x_6$ | 0.803 | -0.123 | 0.064 | 1.935 | 0.10 | 1 |
| $x_3x_4$ | 0.981 | -0.221 | 0.063 | 3.526 | 0.01 | 1 |
| $x_3x_5$ | 0.159 | 0.011 | 0.050 | 0.222 | 0.51 | 0 |
| $x_3x_6$ | 0.786 | 0.130 | 0.075 | 1.746 | 0.89 | 1 |
| $x_4x_5$ | 0 | -0.001 | 0.006 | 0.152 | 0.48 | 0 |
| $x_4x_6$ | 0.998 | -0.158 | 0.021 | 7.341 | 0.00 | 1 |
| $x_5x_6$ | 0.999 | 0.171 | 0.033 | 5.132 | 1.00 | 1 |
| $\sigma_v$ | 1 | 0.048 | 0.002 | 19.67 | 1.00 | 1 |
| $\sigma_u$ | 1 | 0.045 | 0.004 | 10.28 | 1.00 | 1 |

*Notes*: $x_1$ is labor (emp), $x_2$ is private capital stock (pc), $x_3$ is highway and infrastructure capital stock (hwy), $x_4$ is water and sewer facilities capital stock (water), $x_5$ is other public buildings and structures capital stock (util), $x_6$ is state unemployment rate (unempr); the dependent is Gross State Product; see *Produc* dataset documentation in plm package in R; all variables are in natural logs and have been mean-corrected so that first order approximations represent elasticities at the sample mean (i.e., geometric mean). Point estimates of BMA are based on posterior modes.

**Table 6: Correlations of efficiency scores between the *full* model and the *best* model and mean efficiencies**

| | Spearman | Kendall | Mean efficiency full model | Mean efficiency best model |
|---|---|---|---|---|
| Jondrow | 0.9776 | 0.8802 | 0.9548 | 0.9562 |
| Gibbs | 0.9766 | 0.8776 | 0.9525 | 0.9567 |

*Notes*: the best model is the one with the highest posterior probability; see Table 5, column BMS; efficiency comparison is based on two methods: Jondrow et al. (1982) estimator and Gibbs sampling (100 000 draws with initial 10 000 discarded).

**Figure 1: Execution times (runtimes) of the SF-BMA/S procedure with fast pre-screening**

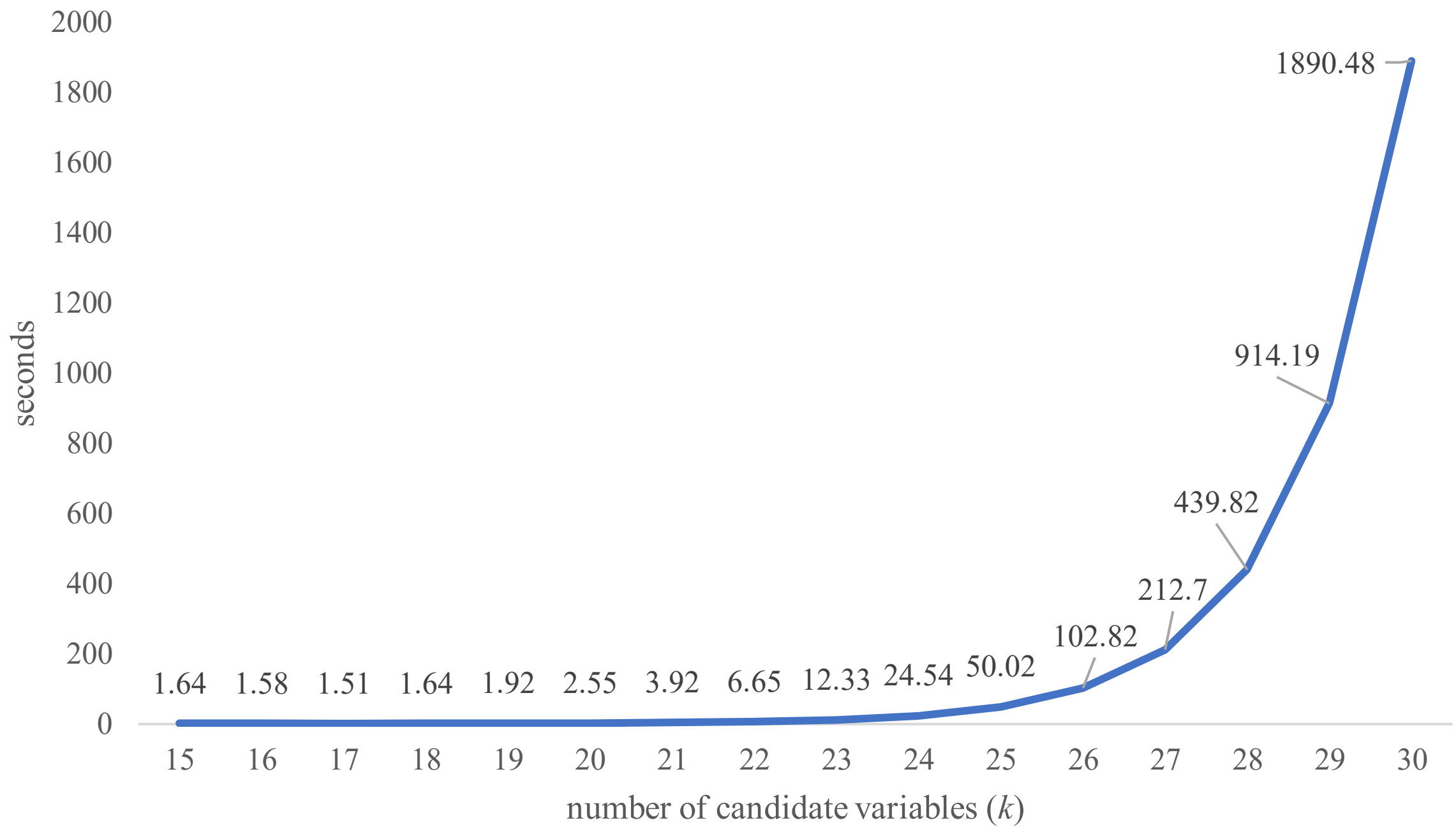

**Figure 2: Marginal posterior distributions of technical efficiency for selected U.S. states**

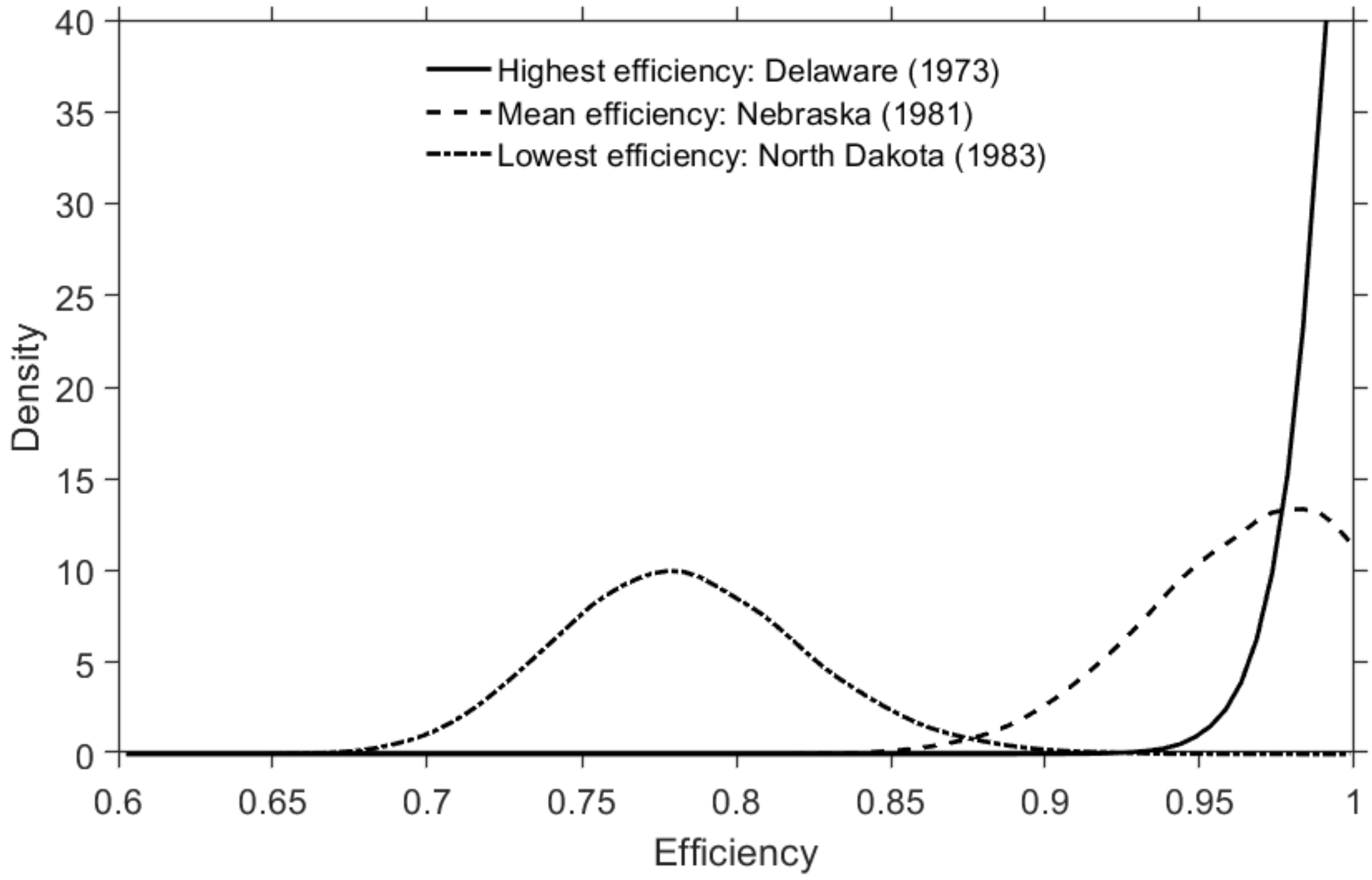

# Appendix: MC results for correlated regressors ($\rho = 0.4$)

**Table A1: Mean differences between accuracy measures in SF-BMA/S and Gaussian BMA/S**

| Sc. | $k_2$ | $\lambda$ | SS | $\overline{\Delta TPR}$ | $-\overline{\Delta FPR}$ | $\Delta ER$ | $\overline{\Delta PIP}_{tr}$ | $-\overline{\Delta PIP}_{fs}$ | $-\overline{-\Delta BS}$ | $-\overline{\Delta RMSE}$ | $\overline{\Delta Sign}$ |
|---|---|---|---|---|---|---|---|---|---|---|---|
| 1 | 4 | 0.5 | 0.25 | 0.005 | -0.005 | -0.010 | 0.002 | -0.001 | 0.001 | 0.002 | 0.003 |
| 2 | 4 | 0.5 | 1 | 0.003 | 0.000 | 0.010 | 0.003 | 0.001 | 0.002 | 0.001 | -0.003 |
| 3 | 4 | 0.5 | 4 | 0.000 | 0.000 | 0.000 | 0.000 | -0.001 | 0.000 | 0.000 | 0 |
| 4 | 4 | 1 | 0.25 | 0.030 | -0.005 | 0.020 | 0.022 | 0.007 | 0.010 | 0.026 | -0.005 |
| 5 | 4 | 1 | 1 | 0.013 | -0.003 | 0.050 | 0.014 | 0.004 | 0.006 | 0.012 | 0 |
| 6 | 4 | 1 | 4 | 0.000 | 0.003 | 0.010 | 0.004 | 0.003 | 0.001 | 0.006 | 0 |
| 7 | 4 | 4 | 0.25 | 0.155 | 0.010 | 0.250 | 0.150 | 0.035 | 0.061 | 0.178 | 0.01 |
| 8 | 4 | 4 | 1 | 0.143 | 0.003 | 0.540 | 0.132 | 0.020 | 0.053 | 0.093 | 0 |
| 9 | 4 | 4 | 4 | 0.010 | 0.008 | 0.060 | 0.017 | 0.012 | 0.007 | 0.039 | 0 |
| 10 | 8 | 0.5 | 0.25 | 0.010 | 0.001 | 0.000 | 0.005 | 0.001 | 0.001 | 0.006 | 0 |
| 11 | 8 | 0.5 | 1 | 0.005 | 0.000 | 0.020 | 0.001 | 0.000 | 0.000 | 0.001 | 0 |
| 12 | 8 | 0.5 | 4 | 0.003 | 0.000 | 0.010 | 0.001 | 0.000 | 0.000 | 0.000 | 0 |
| 13 | 8 | 1 | 0.25 | 0.038 | 0.006 | 0.000 | 0.031 | 0.008 | 0.010 | 0.031 | 0 |
| 14 | 8 | 1 | 1 | 0.005 | -0.003 | 0.010 | 0.010 | 0.002 | 0.003 | 0.007 | -0.003 |
| 15 | 8 | 1 | 4 | 0.008 | 0.001 | 0.030 | 0.007 | 0.003 | 0.002 | 0.005 | 0 |
| 16 | 8 | 4 | 0.25 | 0.170 | 0.013 | 0.250 | 0.167 | 0.039 | 0.052 | 0.161 | 0.025 |
| 17 | 8 | 4 | 1 | 0.138 | 0.014 | 0.530 | 0.133 | 0.023 | 0.043 | 0.086 | 0 |
| 18 | 8 | 4 | 4 | 0.008 | 0.001 | 0.040 | 0.011 | 0.008 | 0.002 | 0.025 | 0 |
| 19 | 10 | 0.5 | 0.25 | 0.003 | -0.001 | 0.000 | 0.002 | 0.002 | 0.002 | 0.004 | -0.003 |
| 20 | 10 | 0.5 | 1 | 0.005 | 0.001 | 0.010 | 0.002 | 0.001 | 0.001 | 0.002 | -0.003 |
| 21 | 10 | 0.5 | 4 | -0.003 | -0.002 | -0.030 | 0.000 | 0.000 | 0.000 | 0.000 | 0 |
| 22 | 10 | 1 | 0.25 | 0.018 | -0.001 | 0.050 | 0.022 | 0.005 | 0.005 | 0.016 | -0.008 |
| 23 | 10 | 1 | 1 | 0.023 | 0.004 | 0.100 | 0.018 | 0.005 | 0.005 | 0.010 | 0.003 |
| 24 | 10 | 1 | 4 | 0.003 | 0.001 | 0.010 | 0.006 | 0.001 | 0.001 | 0.006 | 0 |
| 25 | 10 | 4 | 0.25 | 0.178 | 0.007 | 0.260 | 0.175 | 0.037 | 0.048 | 0.146 | 0.038 |
| 26 | 10 | 4 | 1 | 0.130 | -0.003 | 0.460 | 0.124 | 0.014 | 0.030 | 0.073 | 0 |
| 27 | 10 | 4 | 4 | 0.003 | 0.007 | 0.080 | 0.006 | 0.014 | 0.004 | 0.026 | 0 |
| | | | **Overall** | **0.0406** | **0.0021** | **0.1022** | **0.0394** | **0.0090** | **0.0128** | **0.0357** | **0.0020** |

**Table A2: Scores between accuracy measures in SF-BMA/S and Gaussian BMA/S**

| Sc. | $k_2$ | $\lambda$ | SS | $\overline{\Delta TPR}$ | $-\overline{\Delta FPR}$ | $\Delta ER$ | $\overline{\Delta PIP}_{tr}$ | $-\overline{\Delta PIP}_{fs}$ | $-\overline{\Delta BS}$ | $-\overline{\Delta RMSE}$ | $\overline{\Delta Sign}$ |
|---|---|---|---|---|---|---|---|---|---|---|---|
| 1 | 4 | 0.5 | 0.25 | 0.51 | 0.49 | 0.50 | 0.53 | 0.57 | 0.57 | 0.54 | 0.51 |
| 2 | 4 | 0.5 | 1 | 0.51 | 0.50 | 0.51 | 0.57 | 0.52 | 0.59 | 0.56 | 0.50 |
| 3 | 4 | 0.5 | 4 | 0.50 | 0.50 | 0.50 | 0.61 | 0.55 | 0.57 | 0.50 | 0.50 |
| 4 | 4 | 1 | 0.25 | 0.56 | 0.49 | 0.51 | 0.72 | 0.64 | 0.68 | 0.64 | 0.49 |
| 5 | 4 | 1 | 1 | 0.53 | 0.50 | 0.53 | 0.67 | 0.68 | 0.68 | 0.69 | 0.50 |
| 6 | 4 | 1 | 4 | 0.50 | 0.51 | 0.51 | 0.81 | 0.64 | 0.68 | 0.68 | 0.50 |
| 7 | 4 | 4 | 0.25 | 0.77 | 0.52 | 0.63 | 0.97 | 0.88 | 0.87 | 0.94 | 0.52 |
| 8 | 4 | 4 | 1 | 0.78 | 0.51 | 0.77 | 0.97 | 0.86 | 0.96 | 0.92 | 0.50 |
| 9 | 4 | 4 | 4 | 0.52 | 0.52 | 0.53 | 1.00 | 0.90 | 0.88 | 0.88 | 0.50 |
| 10 | 8 | 0.5 | 0.25 | 0.52 | 0.51 | 0.50 | 0.60 | 0.56 | 0.65 | 0.63 | 0.50 |
| 11 | 8 | 0.5 | 1 | 0.51 | 0.50 | 0.51 | 0.53 | 0.57 | 0.51 | 0.61 | 0.50 |
| 12 | 8 | 0.5 | 4 | 0.51 | 0.50 | 0.51 | 0.66 | 0.57 | 0.55 | 0.59 | 0.50 |
| 13 | 8 | 1 | 0.25 | 0.58 | 0.53 | 0.50 | 0.75 | 0.71 | 0.67 | 0.73 | 0.50 |
| 14 | 8 | 1 | 1 | 0.51 | 0.49 | 0.51 | 0.66 | 0.68 | 0.63 | 0.63 | 0.50 |
| 15 | 8 | 1 | 4 | 0.52 | 0.51 | 0.52 | 0.80 | 0.66 | 0.67 | 0.63 | 0.50 |
| 16 | 8 | 4 | 0.25 | 0.80 | 0.55 | 0.63 | 0.98 | 0.87 | 0.94 | 0.98 | 0.55 |
| 17 | 8 | 4 | 1 | 0.78 | 0.55 | 0.77 | 0.99 | 0.84 | 0.92 | 0.96 | 0.50 |
| 18 | 8 | 4 | 4 | 0.52 | 0.51 | 0.52 | 1.00 | 0.81 | 0.77 | 0.89 | 0.50 |
| 19 | 10 | 0.5 | 0.25 | 0.51 | 0.50 | 0.50 | 0.51 | 0.59 | 0.60 | 0.58 | 0.50 |
| 20 | 10 | 0.5 | 1 | 0.51 | 0.51 | 0.51 | 0.49 | 0.57 | 0.56 | 0.54 | 0.50 |
| 21 | 10 | 0.5 | 4 | 0.50 | 0.49 | 0.49 | 0.51 | 0.59 | 0.60 | 0.53 | 0.50 |
| 22 | 10 | 1 | 0.25 | 0.54 | 0.49 | 0.53 | 0.73 | 0.73 | 0.68 | 0.60 | 0.49 |
| 23 | 10 | 1 | 1 | 0.55 | 0.52 | 0.55 | 0.67 | 0.73 | 0.71 | 0.64 | 0.51 |
| 24 | 10 | 1 | 4 | 0.51 | 0.51 | 0.51 | 0.74 | 0.66 | 0.68 | 0.65 | 0.50 |
| 25 | 10 | 4 | 0.25 | 0.79 | 0.54 | 0.63 | 0.98 | 0.85 | 0.93 | 0.94 | 0.58 |
| 26 | 10 | 4 | 1 | 0.76 | 0.49 | 0.73 | 0.97 | 0.86 | 0.94 | 0.97 | 0.50 |
| 27 | 10 | 4 | 4 | 0.51 | 0.54 | 0.54 | 1.00 | 0.90 | 0.83 | 0.93 | 0.50 |
| | | | **Overall** | **0.575** | **0.508** | **0.551** | **0.756** | **0.703** | **0.716** | **0.718** | **0.504** |